\documentclass[aps, prb, twocolumn, amssymb, amsmath, showpacs, superscriptaddress]{revtex4-1}

\usepackage{bm}
\usepackage{times}
\usepackage{graphicx}
\usepackage{color}
\usepackage{dcolumn}
\usepackage[colorlinks=true, letterpaper=true, pdfstartview=FitV, linkcolor=blue, citecolor=blue, urlcolor=blue]{hyperref}
\usepackage{appendix}
\usepackage[normalem]{ulem}

\begin{document}
\title{Gapless superconducting state and mirage gap in altermagnets}
\author{Miaomiao Wei}
\affiliation{College of Physics and Optoelectronic Engineering, Shenzhen University, Shenzhen 518060, China}
\author{Longjun Xiang}
\affiliation{College of Physics and Optoelectronic Engineering, Shenzhen University, Shenzhen 518060, China}
\author{Fuming Xu}
\affiliation{College of Physics and Optoelectronic Engineering, Shenzhen University, Shenzhen 518060, China}
\affiliation{Quantum Science Center of Guangdong-Hongkong-Macao Greater Bay Area (Guangdong), Shenzhen 518045, China}
\author{Lei Zhang}
\email[]{zhanglei@sxu.edu.cn}
\affiliation{State Key Laboratory of Quantum Optics and Quantum Optics Devices, Institute of Laser Spectroscopy,
Shanxi University, Taiyuan 030006, China}
\affiliation{Collaborative Innovation Center of Extreme Optics, Shanxi University, Taiyuan 030006, China}
\author{Gaomin Tang}
\email[]{gmtang@gscaep.ac.cn}
\affiliation{Graduate School of China Academy of Engineering Physics, Beijing 100193, China}
\author{Jian Wang}
\email[]{jianwang@hku.hk}
\affiliation{College of Physics and Optoelectronic Engineering, Shenzhen University, Shenzhen 518060, China}
\affiliation{Department of Physics, The University of Hong Kong, Pokfulam Road, Hong Kong, China}
\affiliation{Department of Physics, University of Science and Technology of China, Hefei, Anhui 230026, China}

\begin{abstract}

The interplay between spin-orbit interaction (SOI) and magnetism produces interesting phenomena in superconductors. When a two-dimensional (2D) system with strong SOI is coupled to an $s$-wave superconductor, an in-plane magnetic field can drive the system into a gapless superconducting state and induce a mirage gap at finite energies for an Ising superconductor. In this work, we demonstrate that when an $s$-wave superconductor is proximitized to an altermagnet, the intrinsic anisotropic spin splitting of the altermagnet can result in a gapless superconducting state and a pair of mirage gaps at finite energy. The gapless superconductivity exhibits spin-polarized segmented Fermi surfaces, with coexisting spin-singlet and spin-triplet pairings that have a $d$-wave character. Importantly, the gapless superconducting and mirage gap features are quantified through quantum transport. Our results suggest that altermagnet is an ideal platform for studying gapless superconducting states and mirage gap physics.

\end{abstract}

\maketitle

\noindent{\it Introduction} --- The interplay of magnetism and superconductivity is an important research arena in condensed matter physics\cite{RMP1, RMP2, RMP3}. While magnetism can hinder conventional superconducting pairing, causing superconductivity to cease when the magnetic field exceeds the Pauli limit\cite{Clogston}, it can also enable unconventional superconductivity with finite momentum and/or triplet pairing, leading to intriguing physics. For example, in a 2D system with strong spin-orbit interaction (SOI) proximity coupled with an $s$-wave superconductor, an in-plane magnetic field can partially disrupt pairing. This results in a segmented Fermi surface that can be used to create Majorana bound states, provide insights into spin textures of the electron Fermi surface in the normal state and characterize the Fulde-Ferrell-Larkin-Ovchinnikov state in unconventional superconductors\cite{L-Fu1, L-Fu3, L-Fu2}. Experimental evidence of this gapless superconducting state has recently been observed\cite{L-Fu2}. Note that, in addition to the discussed gapless superconducting states, there are two other types: one featuring a Bogoliubov Fermi surface, where the gapless states arise from the form factor in the excitation spectrum, as seen in $p$-wave and $d$-wave superconductors\cite{Agterberg, Brydon, Sim}; and the other created by applying an external magnetic field (below the Pauli limit) to the superconductor, where the gap is fully closed along the entire Fermi surface, as observed in Ref.~[\onlinecite{J-Lee}]. Furthermore, in an Ising superconductor\cite{KT-Law1, J-Lu, X-Xi, Y-Saito, QF-Sun1, QF-Sun2} with an in-plane magnetic field, the presence of equal-spin triplet pairing at finite energy results in a mirage gap that coexists with the quasi-particle density of states\cite{GM-Tang, GM-Tang1}.



Recently, in addition to the ferromagnetic phase and antiferromagnetic phase, a third magnetic phase called the altermagnetic phase has been identified\cite{AM0, AM1, AM2, AM3, AM4, AM5, C-Song1, Nitta, C-Song2}. The altermagnet (AM) has a collinear antiferromagnetic structure with a large non-relativistic anisotropic spin splitting (ASS)\cite{Hayami1,Hayami2}, which leads to several interesting physics phenomena unique to AM. These include giant and tunneling magnetoresistance\cite{AM1}, anomalous spin Hall effect\cite{AM4, AM6, AM7, NC1}, spin splitting torque, and T-odd spin Hall effect\cite{AM5, C-Song1, Nitta, C-Song2}, pronounced thermal transport\cite{Y-Yao1}, and the spin Seebeck and spin Nernst effect of magnons in the absence of Berry curvature due to the giant spin splitting of the magnonic band\cite{Sinova1, T-Yu}. Furthermore, there are numerous materials that exhibit the AM phase, such as RuO$_2$, MnTe, CrO, and CrSb, spanning from insulators, semiconductors, semimetals to metallic systems\cite{AM3}, making it an ideal platform for material engineering\cite{Q-Liu, Y-Yao, Lovesey, Sattigeri}.

When an AM is sandwiched between two superconducting leads, $0$-$\pi$ oscillation is predicted due to the finite momentum pairing\cite{Brataas1, SB-Zhang, Beenakker}. The dependence of Andreev reflection on the orientation of AM relative to the interface, impurity disorder, and tunneling barrier has been studied\cite{Papaj, Brataas2}. Additionally, it has been shown that first and second order topological superconductivity can emerge in 2D AM metals\cite{D-Zhu,Hughes}.

In order to achieve a gapless superconducting state and mirage gap in a 2D system proximitized to an $s$-wave superconductor, it is typically necessary to have an in-plane magnetic field and effective SOI. However, the in-plane magnetic field can potentially destroy the proximitized superconducting state before the gapless superconducting state is created, resulting in a very narrow window for control and manipulation of the gapless state. Our work demonstrates that the use of an in-plane magnetic field is not necessary. Instead, by tuning ASS, the AM proximitized to an $s$-wave superconductor (AM-SC) can transform the $s$-wave superconductor into a gapless superconductor with a segmented Fermi surface resembling a $d$-wave pattern. This segmented Fermi surface exhibits different spin polarizations in different $k$-directions. Additionally, when the quasi-particle energy is nonzero, both singlet and triplet pairing are allowed. Simultaneously, a mirage gap also appears due to finite energy pairing. Finally, these compelling features are identified by the quantum transport calculations.

\begin{figure}
\centering
\includegraphics[width=0.85\columnwidth]{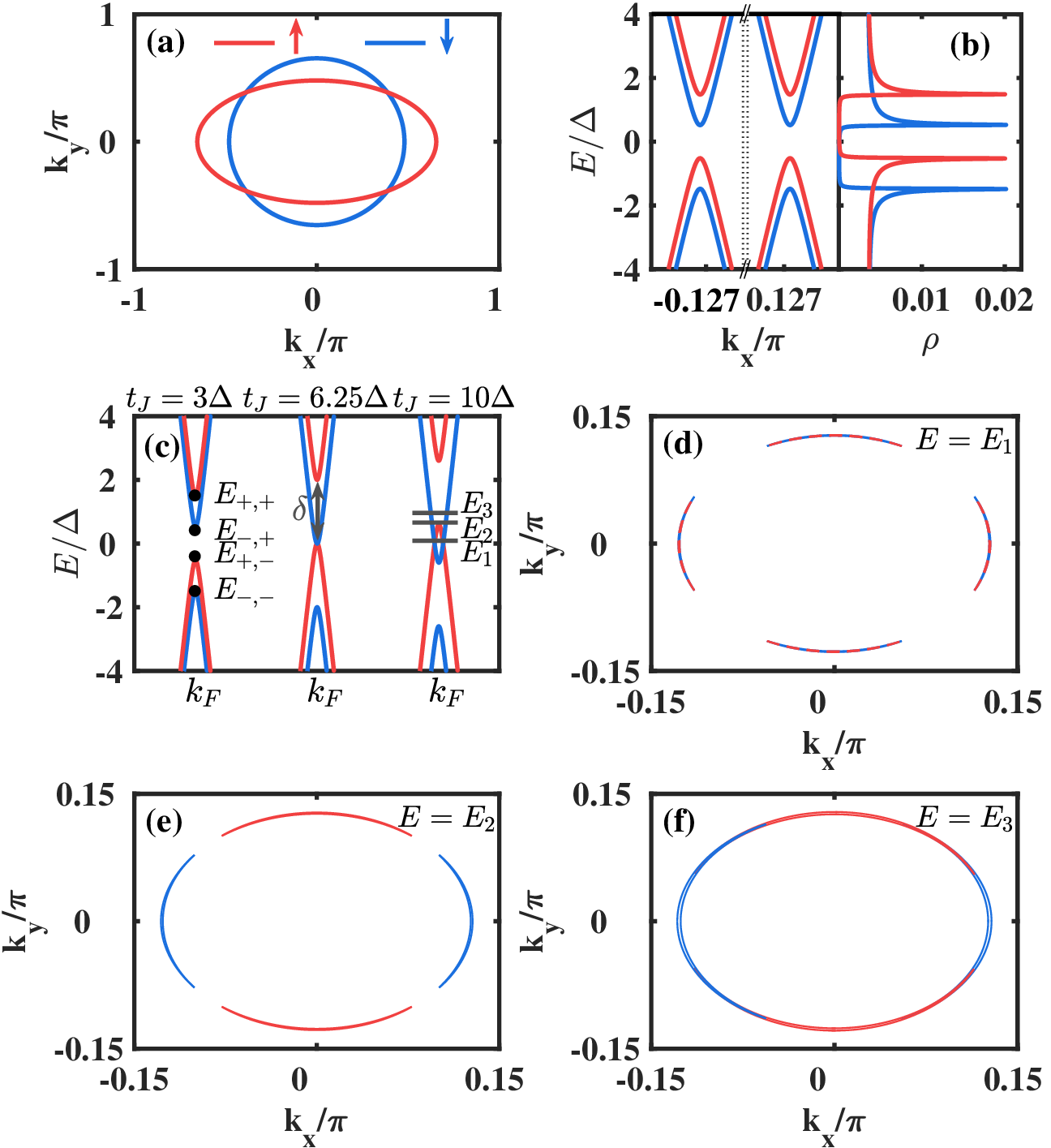}
\caption{(a) Schematic Fermi surfaces in the normal states of the AM. (b) Quasi-particle spectrum of Eq. (\ref{ham}) and corresponding density of states (DOS) with $t_J=3\Delta$. (c) Quasi-particle energy spectrum near the Fermi momentum $k_F$ under different $t_J$ values when $k_y=0$. (d-f) The Fermi surfaces for ${E_1}=0$, ${E_2}=0.6\Delta$, and ${E_3}=2\Delta$ with $t_J=10\Delta$.
The red solid lines represent the spin-up components, while the blue solid lines represent the spin-down components. Note that in panel (d), when $E=0$, the spin is completely degenerate. The spin-up and spin-down components are distinguished by solid and dashed lines, respectively.} \label{FIG1}
\end{figure}

\bigskip

\noindent{\it Quasi-particle spectrum} ---
The Hamiltonian of AM is given by
\begin{eqnarray}
H_0(\textbf{k}) = {{\xi }_{k}}\sigma_0 + {t_J}(k_x^2 -k_y^2)\sigma_z  \nonumber
\end{eqnarray}
where ${{\xi }_{k}}= ( t k^{2} - \mu), k^{2}=k^2_x+k^2_y$, $\mu$ is the chemical potential, and $\sigma_i$ is the Pauli matrix in spin space. ${t_J}$ denotes the strength of ASS in the altermagnet. 
The Fermi momentum is defined as ${{k}_{F}}=\sqrt{{\mu }/{t}\;}$. The spin-splitting Fermi surfaces in the AM are highly anisotropic, as shown in Fig. \ref{FIG1}(a). The spin-up (red solid line) and spin-down (blue solid line) bands are well separated in the $k_x$ and $k_y$ directions, with the spin splitting having opposite signs. However, there is still degeneracy along the directions of $|{k_x}|=|{k_y}|$ due to the $C_4$ symmetry. This type of ASS $d$-wave-like Fermi surface is referred to as $d$-wave magnetism\cite{AM3}.

When the altermagnet is proximitized by an $s$-wave superconductor, the Bogoliubov-de Gennes (BdG) Hamiltonian in the Nambu basis $( {{c}_{\textbf{k},\uparrow }},{{c}_{\textbf{k},\downarrow }},c_{-\textbf{k},\uparrow }^{\dagger },c_{-\textbf{k},\downarrow }^{\dagger } )$ becomes\cite{SC-Zhang}
\begin{equation}
H_{BdG} = \left( \begin{matrix}
   {{H}_{0}}( \textbf{k} ) & \Delta i{{\sigma }_{y}}  \\
   -\Delta i{{\sigma }_{y}} & -H_{0}^{*}( -\textbf{k} )  \\
\end{matrix} \right),
\label{ham}
\end{equation}
where $\Delta$ represents the proximity-induced pairing potential. The spin of Bogoliubov quasi-particles is defined as ${{\mathbf{S}}_{k}}=\langle c_{k}^{\dagger }\boldsymbol{\sigma}{{c}_{k}}-h_{k}^{\dagger }{{\boldsymbol{\sigma}}^{*}}{{h}_{k}} \rangle$\cite{L-Fu1}, where $c_{k}=( c_{k,\uparrow },c_{k,\downarrow } )$ and ${{h}_{k}}=( c_{-k,\uparrow }^{\dagger },c_{-k,\downarrow }^{\dagger } )$ are the electron annihilation and hole creation operators at momentum $k$, respectively, and $\langle {\cdots} \rangle$ denotes average over the eigenstates of the BdG Hamiltonian. Note that $H_{BdG}$ commutes with the spin operator ${{S}_{z}}$.\cite{note2} This indicates that the quasi-particle spectrum exhibits spin-splitting characteristics. In the calculation, we set $t=1$, $\mu=0.16$, and $\Delta=0.001$\cite{unit}.

By diagonalizing the BdG Hamiltonian, one can determine the quasi-particle spectrum. Figure \ref{FIG1}(b) shows the quasi-particle spectrum as a function of $k_x$, with $k_y$ fixed at 0. It is observed that there is a main superconducting gap for both spin components. Additionally, by calculating the spin-dependent density of states (DOS) in Fig. \ref{FIG1}(b), we notice that the superconducting gaps experience spin splitting due to the presence of the exchange interaction $t_J$. In reality, numerous AM materials exhibit different exchange interactions\cite{AM2}. Hence, our objective is to comprehend the superconducting properties of these materials when they are in proximity to an $s$-wave superconductor.

To analytically analyze the evolution of the quasi-particle spectrum versus $t_J$, our focus is on the Fermi momentum $k_F$. After diagonalizing the BdG Hamiltonian, four eigenvalues $E_{\pm,\pm}=\pm T_{J} \pm \Delta$, where $T_J= t_J k_{F}^{2} \cos2\theta, \theta =\arctan ({{{k}_{F,y}}}/{{{k}_{F,x}}})$ and $k_{F,x/y}$ represents the Fermi momenta along the $k_x/k_y$ direction. The ASS results in a direction-dependent spin splitting, which is maximum at $\theta =n{\pi }/{2}, n=0,1,2,3$ and zero for $\theta =n{\pi }/{4}, n=1,3,5,7$. The main superconducting gap is determined by $E_{-+} - E_{+-} = (2\Delta - 2T_J)$. When $t_J$ is zero, there is no spin-splitting superconducting gap, i.e., $E_{+/-,+}-E_{+/-,-}=2\Delta$. For small $T_J<\Delta$, we have $E_{-,-} < E_{+,-} < E_{-,+} < E_{+,+}$. As $t_J$ increases, the main superconducting gap $(2\Delta - 2T_J)$ decreases and eventually closes when $T_J =\Delta$ for some $k$ points. Since the spin splitting vanishes when $\theta =n{\pi }/{4}, n=1,3,5,7$ due to $C_4$ symmetry, the main superconducting gap always remains open at these specific $k$ points. At the same time, a finite energy superconducting gap can be formed. This type of finite energy superconducting gap is referred to as a mirage gap \cite{GM-Tang}. The location (mid point of the gap) and width of the mirage gap are determined by
\begin{equation}
\begin{split}
{{\varepsilon }^{\pm}_{0}}&={\left( {{E}_{\pm,+}}+{{E}_{\pm,-}} \right)}/{2}\;=\pm t_J k_F^2 cos2\theta,\\
\delta &={{E}_{+,+}}-{{E}_{+,-}}={{E}_{-,+}}-{{E}_{-,-}}=2\Delta.
\label{width}
\end{split}
\end{equation}
As $t_J$ increases further, the order of the eigenvalues can change. Specifically, $E_{-,-} < E_{-,+} < E_{+,-} < E_{+,+}$. This implies that normal states appear at $E=0$, and a segmented Fermi surface emerges in the system.

\begin{figure}[ht!]
\centering
\includegraphics[width=0.80\columnwidth]{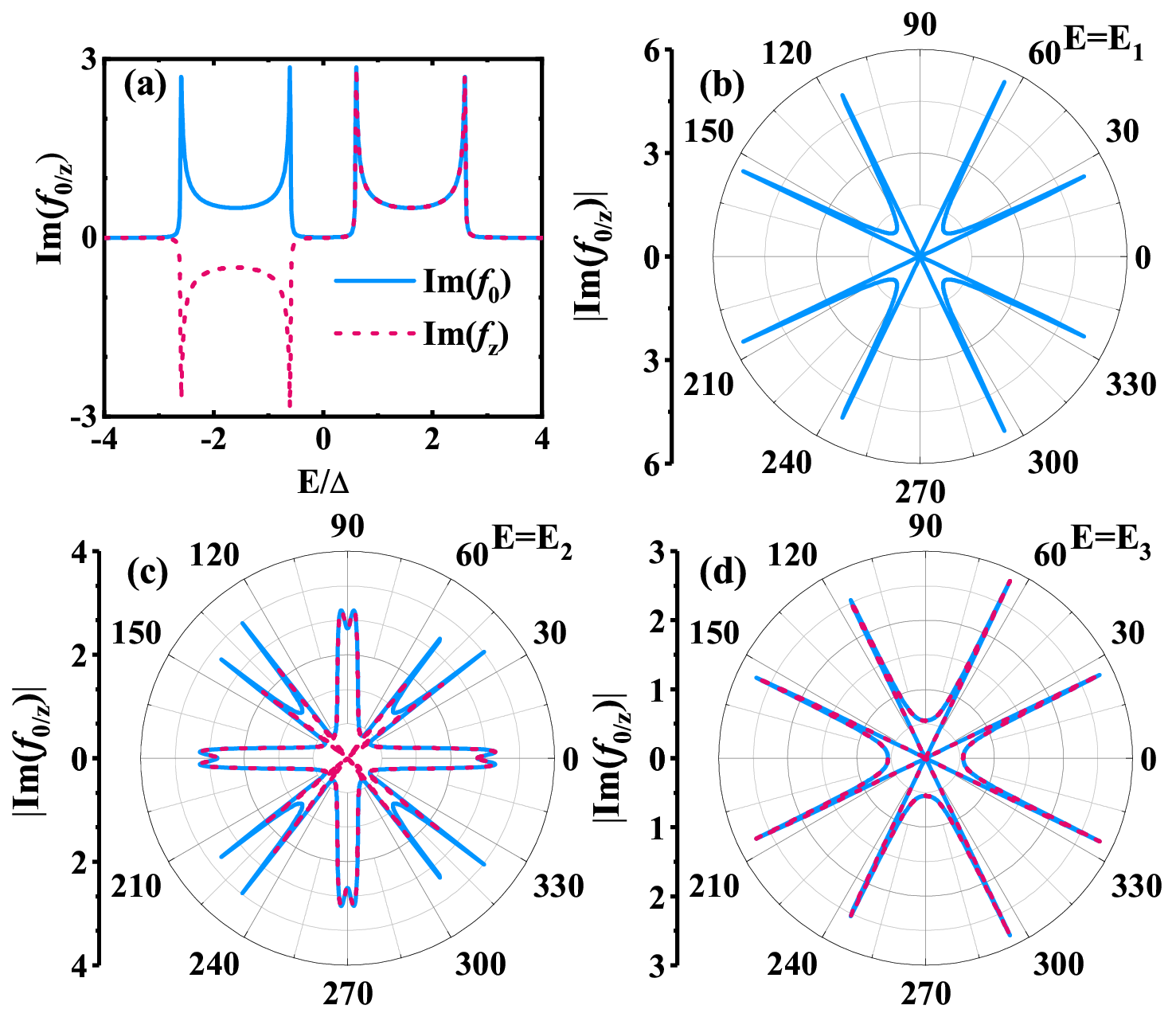}
\caption{
(a) The pair correlations $f_0$ and $f_z$ as a function of $E$ for $t_J=10\Delta$ when $k_{F,y}=0$. (b-d) Polar plot of the pair correlations at different energies $E_1=0$, $E_2=0.6\Delta$, and $E_3=2\Delta$ for $t_J=10\Delta$.}
\label{FIG2}
\end{figure}
Figure \ref{FIG1} (c) shows a plot of the quasi-particle spectrum evolution under different ASS strengths $t_J = 3\Delta, 6.25\Delta$ and $10\Delta$ with $k_{y}=0$. Notably, the main superconducting gap diminishes and closes as $t_J$ increases. When it reaches a critical value of $6.25\Delta$, the main superconducting gap closes and a mirage gap $\delta$ forms, as depicted in the middle panel of Fig. \ref{FIG1} (c). Furthermore, upon further increasing $t_J=10\Delta$, although the main superconducting gap disappears at $E=0$, a pair of mirage gaps still exist at finite energy ($|{\varepsilon }_{0}|\geq \Delta$).
In this situation, we investigate three different Fermi surfaces, which are presented in Figs. \ref{FIG1} (d-f).
Due to the vanishing of spin splitting at $\theta =n{\pi }/{4}, n=1,3,5,7$, segmented Fermi surfaces are formed at $|E|<\Delta$, and the superconducting gap always appears around these specific points. The segmented Fermi surfaces indicate the presence of gapless superconducting states in the system. Interestingly, there is no spin splitting for the segmented Fermi surfaces when $E=0$. As the energy increases, the spin splitting starts to appear and eventually fully separates in $k$-space. For instance, when $E=E_2$, we observe spin-polarized segmented Fermi surfaces in Fig. \ref{FIG1}(e). Here, the main superconducting gap and mirage gaps coexist. Even when the energy exceeds $\Delta$, the mirage gap persists. There is some overlap between the spin-up and spin-down surfaces, as depicted in Fig. \ref{FIG1} (f). Therefore, AM materials offer an ideal platform for observing both gapless superconducting and mirage gap physics simultaneously. In the next section, we will explore the pairing mechanism of the superconducting gaps.


\noindent{\it Pairing mechanism} ---
The general pairing correlation function is defined as follows:\cite{Gorkov, Sigrist, GM-Tang}
\begin{eqnarray}
{\cal F}({\bf k}, E) = \Delta (F_0 \sigma_0 + {\bf F} \cdot {\boldsymbol{\sigma}})  i\sigma_y,
\label{F0}
\end{eqnarray}
where $F_0$ and ${\bf F}$ represent the singlet and triplet pairing correlations, respectively. We can use ${\cal F}({\bf k}, E)$ to express the pairing wave function
\begin{equation}
\begin{split}
\left| \Psi  \right\rangle =&{{F}_{0}}\left( \left| \uparrow \downarrow  \right\rangle -\left| \downarrow \uparrow  \right\rangle  \right)+{{F}_{x}}\left( \left| \downarrow \downarrow  \right\rangle -\left| \uparrow \uparrow  \right\rangle  \right)\\
&+i{{F}_{y}}\left( \left| \downarrow \downarrow  \right\rangle +\left| \uparrow \uparrow  \right\rangle  \right)+{{F}_{z}}\left( \left| \uparrow \downarrow  \right\rangle +\left| \downarrow \uparrow  \right\rangle  \right).
\end{split}
\end{equation}
In the above expressions, the abbreviations $| \textbf{k}\uparrow ,-\textbf{k}\downarrow \rangle$ are used to represent $\mid \uparrow \downarrow \rangle$. The pairing correlations can be obtained by solving the Gorkov equation,\cite{Gorkov, Sigrist, book1, GM-Tang}
\begin{equation}
\left[ \begin{matrix}
   E -{{H}_{0}}( \textbf{k} ) & -\Delta i{{\sigma }_{y}}  \\
   \Delta i{{\sigma }_{y}} & E +H_{0}^{*}( -\textbf{k} )  \\
\end{matrix} \right]\left[ \begin{matrix}
   {\cal F}( \textbf{k},E  )  \\
   \bar{G}( \textbf{k},E )  \\
\end{matrix} \right]=\left[ \begin{matrix}
   0  \\
   1  \\
\end{matrix} \right].
\label{F3}
\end{equation}
Eliminating $\bar{G}( \textbf{k},E )$ in Eq. (\ref{F3}),
we have the components of the pairing-correlation function ${\cal{F}}( \textbf{k},E )$ for the proximity coupled altermagnet state
\begin{equation}
\begin{split}
{{F_{0}( \textbf{k},E  )}}&={\left( {{E }^{2}}-{{\Delta }^{2}}-{{\xi }_{k}^{2}} +{t_J^2} \right)}/{M( \textbf{k},E  )}\;,\\
{{F_{z}( \textbf{k},E  )}}&=-{2E t_J}/{M( \textbf{k},E  )}\;,
\label{F2}
\end{split}
\end{equation}
with $M( \textbf{k},E  )= {{( {{E }^{2}}-{{\Delta }^{2}}-{{\xi }_{k}^{2}} +{t_J^2} )}^{2}-4{E^{2}}{t_J^2}}$. From Eq. (\ref{F2}), one can see that, even when in proximity to an $s$-wave superconductor, both spin-singlet and spin-triplet pairings can coexist with a nonzero energy $E$ in AM-SC.
Furthermore, if we focus primarily on phenomena near the Fermi surface, we can introduce the dimensionless Green's functions $f_{0/z}({\bf k}_F, E)$ to characterize the pairing amplitude of $F_{0/z}$\cite{Mockli,Kita}
\begin{equation}
 f_{0/z}({\bf k}_F, E)=\oint{\frac{d{{\xi }_{k}}}{\pi }i{ F}_{0/z}({\bf k}, E)}.
\end{equation}
Fig. \ref{FIG2}(a) presents the dimensionless Green's functions $f_{0/z}({\bf k}_F, E)$ as a function of quasi-particle energy, with $k_{F,y}=0$ and $t_J=10\Delta$. It is observed that there is no pairing formed around $E=0$, indicating that the main superconducting gap is closed at $k_y=0$. However, both singlet and triplet pairings are nonzero in the mirage gap shown in the quasi-particle spectrum in Fig. \ref{FIG1}(c). This also confirms the coexistence of spin-singlet and spin-triplet pairings of AM-SC. Note that $f_0$ and $f_z$ have even and odd parities with respect to the energy $E$. Moreover, we investigate the anisotropic property of pairing by plotting $f_{0/z}({\bf k}_F, E)$ in polar axis under different energies labeled in Fig. \ref{FIG1} (c). From Fig. \ref{FIG2} (b), we observe that the singlet pairing is nonzero around $\theta =n{\pi }/{4}, n=1,3,5,7$. This is consistent with what we found in the quasi-particle spectrum shown in Fig. \ref{FIG1} (d). The AM-SC exhibits a gapless feature. However, the triplet pairing is always zero at $E_1=0$, which is in agreement with Eq. (\ref{F2}). If the quasi-particle energy is increased to $E_2$, triplet pairing emerges, presenting a different feature compared to singlet pairing. When the energy is further shifted upward to $E_3=2\Delta$, the pairings become zero in the spin-overlapping region as shown in Fig. \ref{FIG1} (f). This is different from the pairing at $E_1=0$ in the $k$ space. It is worth noting that the polar plot of the pairings exhibits a four-fold rotational symmetry in all cases. In the following, we will discuss how to use the transport properties to reflect the interesting superconducting gap and pairing information of an AM-SC.

\begin{figure}[ht!]
	\centering
	\includegraphics[width=0.85\columnwidth]{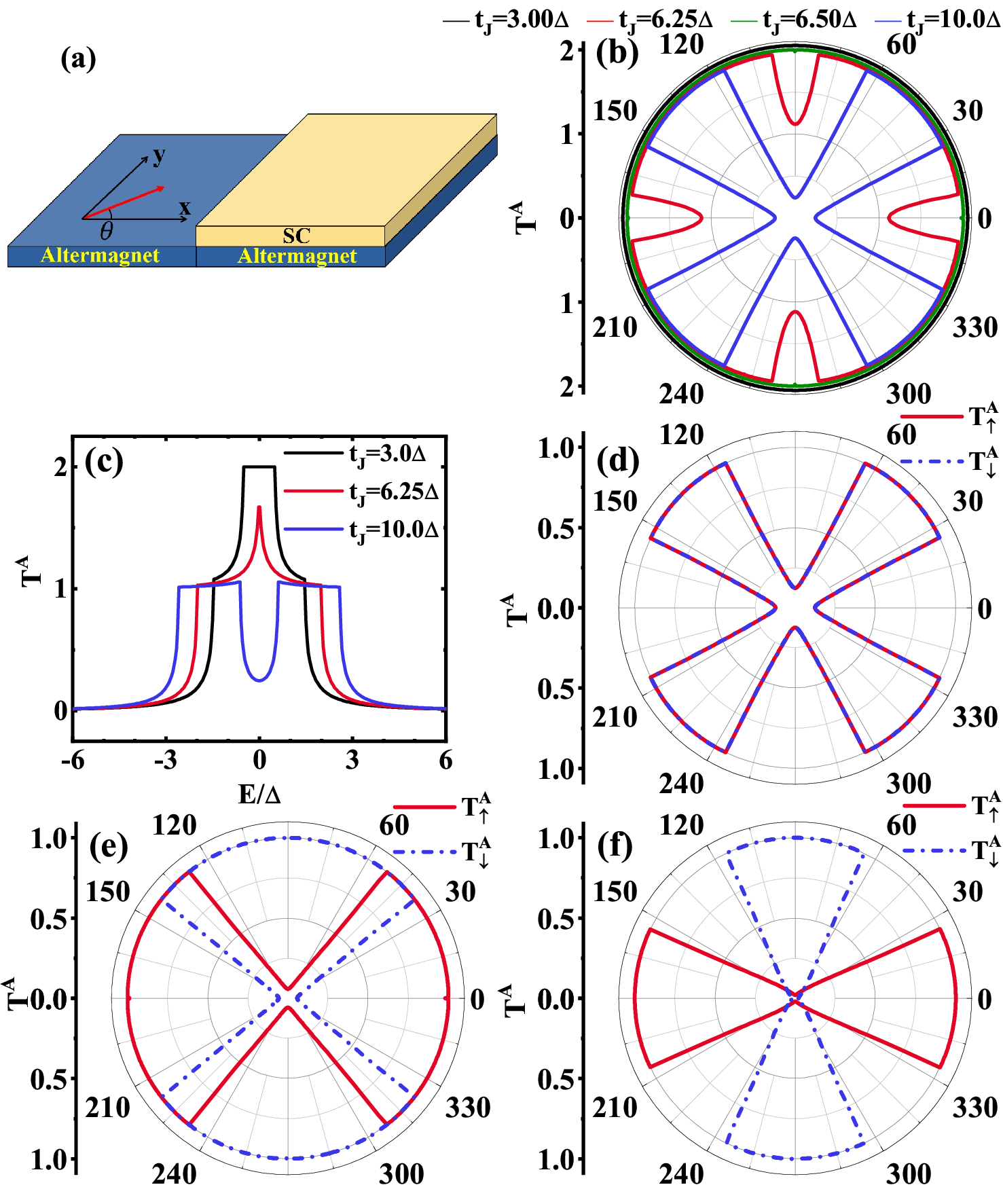}
	\caption{(a) Schematic diagrams of an altermagnet-superconductor junction. $\theta$ represents the incident angle of the electrons in the normal altermagnet. (b) The polar plot of the Andreev reflection coefficients $T^A$ for different $t_J$ values at $E=0$. (c) The Andreev reflection coefficient $T^A$ plotted against $E_F$ for different $t_J$ values with $\theta=0^\circ$. (d-f) Polar plots showing the spin-resolved Andreev reflection coefficients plotted against $\theta$ at different energies: $E=0, 0.6\Delta$, and $2\Delta$ for $t_J=10\Delta$. }\label{FIG3}
\end{figure}

\noindent{\it Quantum transport} --- We primarily study the transport properties (Andreev reflections) of an altermagnet-superconductor junction, as depicted in Fig. \ref{FIG3}(a). In this configuration, the $s$-wave superconductor is positioned on a section of an AM, resulting in Andreev reflection coefficient $T^A$ that can be utilized to identify the associated superconducting information. Detailed quantum transport calculation approaches can be found in the Supplemental Material\cite{note2}. First, the polar plot of the Andreev reflection coefficients $T^A$ changes from isotropic to anisotropic $d$-wave character as the ASS increases when $E=0$, as shown in Fig. \ref{FIG3} (b). The persistence of quantized $T^A$ around $|k_x|=|k_y|$ in all cases corresponds to the disappearance of spin splitting. Fig. \ref{FIG3}(c) shows the Andreev reflection coefficient $T^A$ as a function of $E$ under different ASS strengths $t_J = 3\Delta, 6.25\Delta$ and $10\Delta$, with an electron incident angle of $\theta=0^\circ$. It can be observed that $T^A$ is quantized to $2$ when $|E|<0.48\Delta$ for $t_J=3\Delta$, which indicates the presence of a main superconducting gap for both spin components [see Fig. \ref{FIG1}(c)]. However, this quantization disappears when $t_J$ is increased up to $6.25\Delta$. In this case, the main superconducting gap closes and two spin-splitting mirage gaps at $| {{\varepsilon }_{0}}|=\Delta $ just emerge. Correspondingly, the spin-resolved Andreev reflection coefficients\cite{Melin, YanYangZhang, note1} $T^A_\sigma$ are quantized to 1 within the mirage gap, as shown in Supplemental Material\cite{note2}. Due to the proximity effect, the Andreev reflection coefficients $T^A_\sigma$ do not immediately decrease to zero when the energy is outside the mirage gap, hence a non-quantized value of $T^A$ around $E=0$ in Fig. \ref{FIG3}(c). When $t_J$ is large enough, the two spin-splitting mirage gaps are well separated, and the quantized $T^A$ reappears. However, there are still nonzero Andreev reflection coefficients around $E=0$. This can be seen in Fig. \ref{FIG3}(c), where $T^A$ approaches quantization to $1$ within the mirage gaps for $t_J=10\Delta$. The center of the quantized $T_{\sigma}^A$ corresponds to the location $\varepsilon^{\pm}_0$ of the mirage gap. The range of quantization, which is independent of $t_J$, is equal to $2\Delta$. These results are consistent with the analytic expressions given in Eq. (\ref{width}). Our theoretical calculations yield quantifiable data about the superconducting gap.

\begin{figure}[ht!]
	\centering
	\includegraphics[width=0.85\columnwidth]{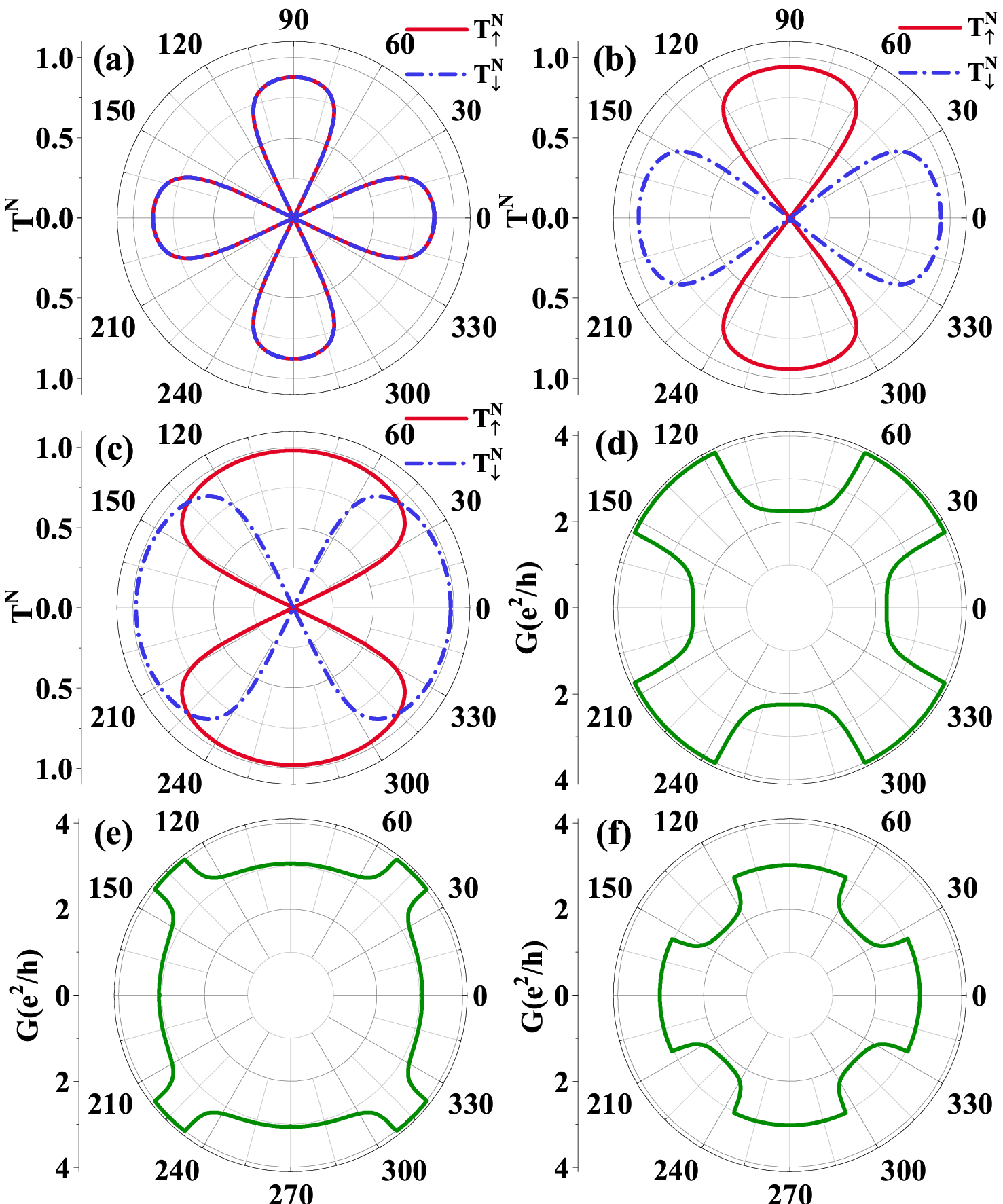}
	\caption{Polar plots showing (a-c) the spin-resolved quasi-particle transmission coefficient $T_{\sigma}^{N}$ and (d-f) the conductance $G$ plotted against $\theta$ at different energies: $E=0, 0.6\Delta$, and $2\Delta$ for $t_J=10\Delta$. }\label{FIG4}
\end{figure}

Next, we will use angle-dependent transport to demonstrate the properties of the segmented Fermi surface in the gapless superconducting phase. In Figs. \ref{FIG3}(d-f), the polar plot displays the spin-resolved Andreev reflection coefficients ($T_{\sigma}^{A}$) versus $\theta$ at three different energies labled in Fig. \ref{FIG1}(c). The red solid lines represent the spin-up Andreev reflection coefficient ($T_{\uparrow}^{A}$), while the blue dashed lines represent the spin-down Andreev reflection coefficient ($T_{\downarrow}^{A}$).
For $E_1=0.0\Delta$, $T_{\uparrow}^{A}$ and $T_{\downarrow}^{A}$ are fully degenerate.
The profile of $T_{\sigma}^{A}$ has $C_4$ symmetry and centers at $\theta =n{\pi }/{4}, n=1,3,5,7$. However, as $E$ increases, they start to separate from each other.
When the Fermi energy increases to $E_2$, the spin-up and spin-down parts connect on the principal horizontal and vertical axes, respectively [see Fig. \ref{FIG3}(e)].
Further increasing $E>\Delta$, i.e., $E_3=2\Delta$ in Fig. \ref{FIG3}(f), both spin components are zero around $\theta =n{\pi }/{4}, n=1,3,5,7$. The profile of $T_{\uparrow}^{A}$ and $T_{\downarrow}^{A}$ reduce to $C_2$ symmetry and rotate to center at the principal horizontal and vertical axes, respectively.
It is observed that the quantized $T_{\sigma}^{A}$ align with the range of the superconducting gap regions shown in Fig. \ref{FIG1}.

To gain comprehensive insight into the full conductance features accessible in experiments, we first present
the polar plot of the quasi-particle transmission coefficients $T_{\sigma}^{N}$, as shown in Fig. \ref{FIG4}(a-c).
The spin-up and spin-down quasi-particle transmission coefficients are represented by red solid and blue dashed lines, respectively. These coefficients, $T_{\uparrow}^{N}$ and $T_{\downarrow}^{N}$, are fully degenerate when $E_1=0.0\Delta$. However, they begin to separate as $E$ increases. When the energy $E_2 = 0.6\Delta$, the spin-up and spin-down parts are well separated, as shown in Fig. \ref{FIG4}(b). Upon further increasing $E>\Delta$ (i.e., $E_3=2\Delta$ in Fig. \ref{FIG4}(c)), both spin components overlap around $\theta =n{\pi }/{4}, n=1,3,5,7$. Furthermore, the quasi-particle transmission coefficient in the polar plot is spin-polarized around $\theta =n{\pi }/{2}, n=0,1,2,3$ when $E\ne 0$.
Comparing with the Andreev reflections, the quasi-particle transmission coefficients $T_{\sigma}^{N}$ align with the segmented Fermi surface.

Finally, we discuss the total conductance, as shown in Fig. \ref{FIG4}(d-f).
In the low bias limit, the conductance in the AM-SC junction can be expressed as $G=\frac{e^2}{h}\sum\nolimits_{\sigma }{( T_{\sigma }^{N}+2T_{\sigma }^{A})}$.
when $E=0$, the conductance reaches $4$ around $\theta =n{\pi }/{4}, n=1,3,5,7$ due to the Andreev
reflections in the main superconducting gap, whereas the conductance dominated by the normal state in other regions are approximately to $2$.
As $E$ increases up to $0.6\Delta$, the conductance in the mirage gap increases up to $3$ around the principal horizontal and vertical axes.
When $E$ equals to $2\Delta$, the total conductance around $\theta =n{\pi }/{4}, n=1,3,5,7$ reduce to $2$ due to the further separation of the spin resolved Andreev reflections. It can be seen that the total conductance keeps $C_4$ symmetry. Given the angle-resolved measurements of transport property of third-order nonlinear Hall effect recently\cite{YangNat16869}, we may anticipate observing the angle-resolved total conductance in the AM-SC system.

\noindent{\it Conclusion} --- In summary, we have demonstrated that in the absence of SOI, the AM-SC can display $d$-wave-like gapless superconducting states. Moreover, both singlet and triplet pairings occur simultaneously at finite energy, resulting in the emergence of mirage gaps in the system. These key features have been validated and quantified through quantum transport calculations. Our findings highlight the AM as an ideal platform for studying gapless superconducting states and mirage gap physics.

\bigskip

\noindent{\it Acknowledgments} ---
This work is supported by the National Natural Science Foundation of China (Grant No. 12034014, 12074230, and 12174262).


\begin{thebibliography}{00}
\bibitem{unit}
$t$ is set to the unit of energy, i.e., $t = 1 eV$.
\end{thebibliography}
\end{document}